\documentclass[twocolumn,twocolappendix]{aastex701}

\usepackage{amsmath}
\usepackage{amssymb}

\newcommand{\rlc}{\ensuremath{R_\mathrm{LC}}}
\newcommand{\rns}{\ensuremath{R_\star}}
\newcommand{\ngj}{\ensuremath{n_\mathrm{GJ}}}
\newcommand{\jgj}{\ensuremath{j_\mathrm{GJ}}}

\newcommand{\Porb}{\ensuremath{P_\mathrm{orb}}}
\newcommand{\Pone}{\ensuremath{P_\mathrm{B'}}}
\newcommand{\Ptwo}{\ensuremath{P_\mathrm{A'}}}
\newcommand{\Omorb}{\ensuremath{\Omega_\mathrm{orb}}}
\newcommand{\vorb}{\ensuremath{v_\mathrm{orb}}}
\newcommand{\Bstar}{\ensuremath{B_\star}}
\newcommand{\gthr}{\ensuremath{\gamma_\mathrm{thr}}}
\newcommand{\gsec}{\ensuremath{\gamma_\mathrm{sec}}}
\newcommand{\Bone}{\ensuremath{B_\mathrm{B'}}}
\newcommand{\Btwo}{\ensuremath{B_\mathrm{A'}}}
\newcommand{\PSR}{PSR~J0737$-$3039A/B}
\newcommand{\starone}{$B'$}
\newcommand{\startwo}{$A'$}

\shorttitle{Kinetic model of an orbiting binary neutron star magnetosphere}
\shortauthors{Ben\'a\v{c}ek et al.}

\begin{document}

\title{Fast Radio Bursts from Monster Shocks and Flares of Inspiraling Binary Neutron Stars:\\ A Kinetic Model for Double Pulsar}

\author[0000-0002-4319-8083]{Jan Ben\'a\v{c}ek}
\affiliation{Astronomical Institute of the Czech Academy of Sciences, Fričova 298, 25165 Ondřejov, The Czech Republic}
\affiliation{Institute for Physics and Astronomy, University of Potsdam, Karl-Liebknecht-Stra\ss{}e 24/25, 14476 Potsdam, Germany}
\email{jan.benacek@example.org}

\author[0000-0001-6295-596X]{Beno\^it Cerutti}
\affiliation{Univ. Grenoble Alpes, CNRS, IPAG, 38000 Grenoble, France}
\email{}

\author[0000-0001-7861-1707]{Martin Pohl}
\affiliation{Institute of Physics and Astronomy, University of Potsdam, Karl-Liebknecht-Stra\ss{}e 24/25, 14476 Potsdam, Germany}
\affiliation{Deutsches Elektronen-Synchrotron DESY, Platanenallee 6, 15738 Zeuthen, Germany}
\email{}

\author[0000-0003-2374-307X]{Tim Dietrich}
\affiliation{Institute for Physics and Astronomy, University of Potsdam, Karl-Liebknecht-Stra\ss{}e 24/25, 14476 Potsdam, Germany}
\affiliation{Max Planck Institute for Gravitational Physics (Albert Einstein Institute), Am M\"uhlenberg 1, Potsdam 14476, Germany}
\email{}

\author[0000-0003-3205-8373]{Anna Neuweiler}
\affiliation{Institute for Physics and Astronomy, University of Potsdam, Karl-Liebknecht-Stra\ss{}e 24/25, 14476 Potsdam, Germany}
\email{}

\begin{abstract}
The coalescence of binary neutron star systems offers 
unprecedented possibilities for multimessenger observations, including the detection of gravitational waves, gamma-ray bursts, kilonovae, and potentially fast radio bursts. 
For a detailed investigation of the latter, one requires a kinetic plasma model to describe the interacting magnetospheres of the stars.
We present a three-dimensional particle-in-cell kinetic simulation of the combined magnetosphere of a binary neutron star in which the two stars physically orbit one another so that the orbital motion itself twists the magnetic loops.
We find that the stellar magnetic-field intensity ratio controls the emergence of flares, while the stellar spins set the direction and modulation of the flare. 
We identify four spatially and dynamically distinct regions of strong electromagnetic energy release: (i) an expanding fast magnetosonic wave that steepens into a monster shock, (ii) a plasma flare being pushed by (iii) an expanding fireball and containing (iv) a reconnecting current sheet. 
The fast magnetosonic wave is a promising source of coherent radio emission.
It is released when orbital motion cuts a closed plasma loop.
The electric and magnetic wave amplitudes are similar and reach that of the background magnetic field.
Our kinetic treatment yields particle spectra and plasma structures that can be used to assess coherent emission in kinetic-scale simulations.
In conclusion, the stellar orbital motion constitutes a driving channel distinct from models with spin-driven magnetospheric twisting.
\end{abstract}


\section{Introduction} \label{sec:intro}
Fast radio bursts (FRBs) are millisecond-duration, extremely luminous radio transients of mostly extragalactic origin \citep{Petroff2019,Cordes2019}. 
Their brightness temperatures reach $\gtrsim10^{35}$~K and require a coherent emission process.
Despite many efforts, the coherent mechanism itself remains unresolved.

One possible source for FRBs is the late inspiral of binary neutron star (BNS) systems.
Due to the emission of gravitational waves from the system, the orbital separation between the two stars decreases. In the final evolutionary phase before the merger, the combined orbital and spin motions strongly twist the magnetosphere, dissipating orbital and magnetic energies. Consequently, an electromagnetic precursor of the merger appears, preceding the gravitational wave signal and subsequent electromagnetic merger counterparts \citep{Palenzuela2013}.
Global plasma magnetospheric models of this phase have predominantly been magnetohydrodynamic and force-free simulations \citep{MostPhilippov2020,MostPhilippov2022,MostPhilippov2023}. 
They showed that twisted flux bundles connecting the stars store energy that is explosively released in periodic, topology-dependent flares that can steepen into shocks and potentially turn into FRBs.
\citet{Carrasco2020} treated a single orbiting NS, and \citet{Skiathas2025} investigated the particle processes and high-energy emission. 

First-principles kinetic modeling of NS magnetospheres focused initially on single-star geometries \citep{PhilippovSpitkovsky2014,Cerutti2015}. 
\citet{Crinquand2019} presented global particle-in-cell (PIC) simulations of pre-merger BNS magnetospheres with stars approaching each other at a merging rate consistent with the emission of gravitational waves. 
Only very recently, \citet{Parsons2026} presented simulations carried out in a frame co-rotating with the binary, in which the eruptions are powered by the stellar spin-induced twisting of the flux bundles connecting the stars, and \citet{Lyutikov2026} considered the orbital motion of a single star, approximated as a linearly moving unipolar inductor.
A kinetic model with the orbital motion as the driver for the magnetospheric dynamics shortly before the merger has not yet been developed.

In this letter, we present a three-dimensional PIC kinetic simulation of the common magnetosphere of a BNS system in which the stars orbit one another in the simulation frame.
This kinetic treatment allows us to investigate the plasma kinetic properties and pair production sites to identify and characterize the candidate coherent emission sites of the inspiraling system inspired by double pulsar observations such as \PSR\, analog, a future potential source of FRBs.
Further detailed analysis of the BNS system emission sources will follow.

\section{Methods and Results} \label{sec:methods}
The binary system model is loosely inspired by the double pulsar \PSR\,
\citep{Burgay2003,Lyne2004,Kramer2006,Kramer2021}, i.e., a system consisting of a primary star with a mass of $1.34~M_\odot$ and a rotation period of $23~\rm ms$, and a secondary star with a mass of $1.25~M_\odot$ and a spin period of $2773~\rm ms$. Due to the emission of gravitational waves, the angular momentum and energy of the double neutron star system are continuously reduced. Minutes before the merger, the stars reach small enough separations and high enough velocities for the emitted gravitational waves to become measurable with ground based gravitational-wave detectors~\citep{LIGO2015,Acernese2015}, as exemplary shown with GW170817~\citep{Abbott2017}. For \PSR\ this will happen in around 85 million years.

We consider the system about 0.1\,s before the merger, i.e., well within the frequency band of currently operating ground-based detectors. Employing the model presented in~\cite{Abac2024}, the binary will have roughly an orbital period of $\Porb=2.4$~ms at this time. In our simulation, the stars are set up to correspond to a classical pulsar, \starone, with a long spin period and a strong magnetic field and a millisecond pulsar, \startwo, with a shorter spin period and a weaker magnetic field than \starone. We let them rotate with spin periods of $\Ptwo=10$~ms and $\Pone=100$~ms. The radii of both stars are set to $\rns=12~\rm km$, in agreement with recent multi-messenger constraints regarding the neutron star equation of state~\citep{Dietrich2020,Koehn2025}. The spin and orbital angular momenta are mutually aligned, which simplifies the setup but does not represent \PSR\ for which the slow-rotating star has an almost antialigned spin. Both stars carry dipolar magnetic fields with the ratio $\Bone=10\,\Btwo$, and both are modeled as orthogonal rotators. We use this idealized configuration to isolate orbital effects.
This setup makes the configuration feasible to run while retaining, as expected for the double pulsar, high magnetic-inclination angles and the scale ordering $\Porb<\Ptwo\ll\Pone$ and $\Bone\gg\Btwo$.

The length scale controlling the magnetospheric dynamics is the orbital light cylinder, $\rlc=c/\Omorb$, where $\Omorb=2\pi/\Porb$ and $c$ is the speed of light. 
For our parameters $\rlc\approx114\,\mathrm{km}\,\approx9.5\rns$.
The light cylinders of both stars are therefore larger than the orbital light cylinder.
Details of the numerical setup are stated in Appendix~\ref{appendixA}.

\begin{figure*}[t]
\centering
\includegraphics[width=0.86\textwidth]{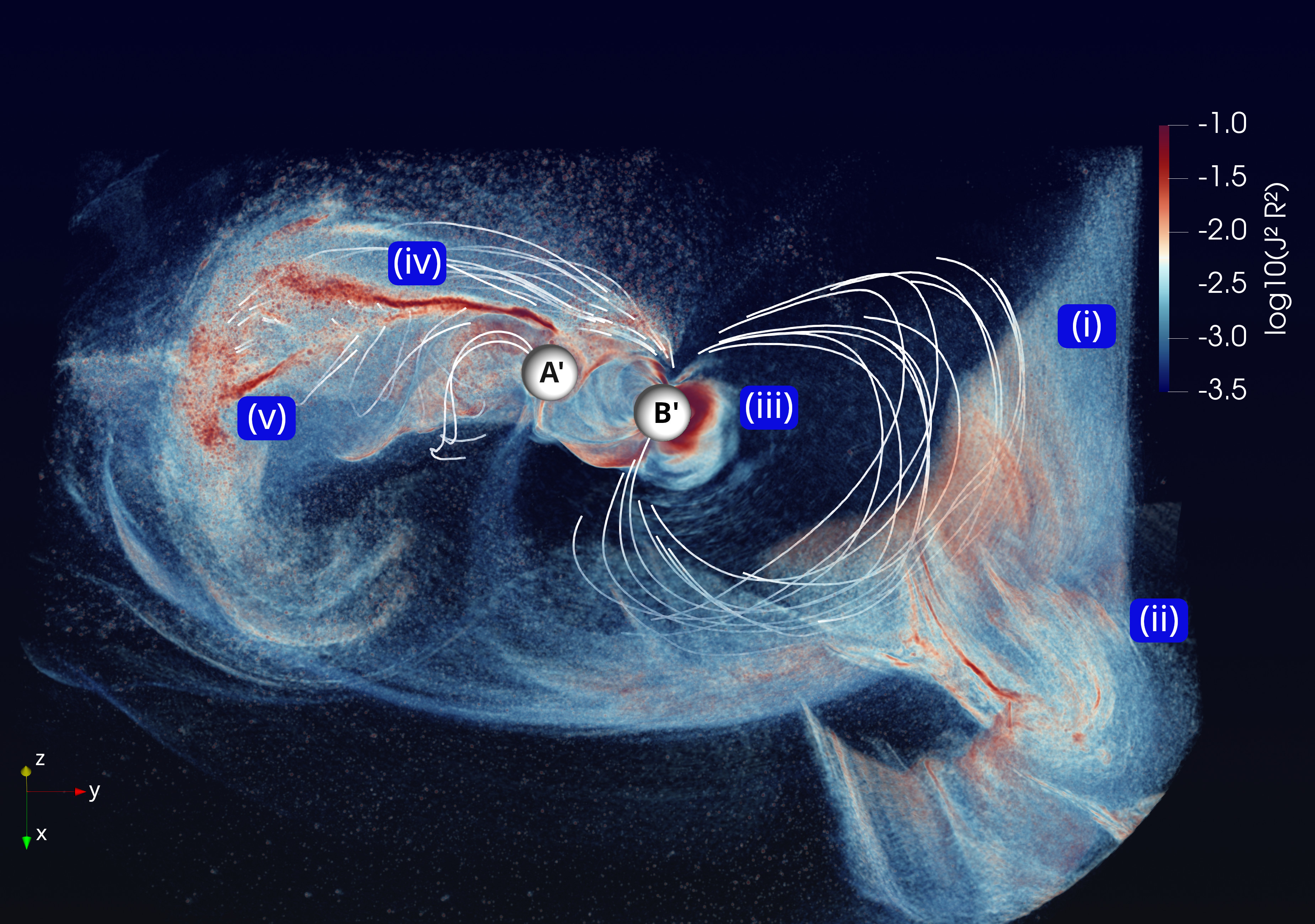}
\caption{
    Three-dimensional rendering of the electric current density $J=|\boldsymbol{j}|/j_\mathrm{GJ}$ below the orbital plane at orbital phase $0.52\Porb$ and scaled with distance from the center of mass $R=|\boldsymbol{r}|/\rlc$.
    White lines trace selected magnetic field lines.
    The two stars are the light gray spheres in the center and rotate clockwise. 
    Labeled features are (i) the expanding FMSW forming the monster shock, (ii) the plasma flare, (iii) the fireball, (iv) the current sheet forming (v) a proto-flare, which is emitted $\approx\Porb/2$ later. 
    \label{fig:render}}
\end{figure*}

\subsection{Appearance of Monster Shock, Flare, and Fireball}\label{sec:orbit}
We started the model development with a simple individual neutron-star magnetosphere and a single star linearly moving through the grid.
We also tested distant and close double NS systems without orbital motion.
The final model we present here reflects the features seen in these tests and is therefore a complex state appearing in the BNS system evolution.
Because of this test ladder, we are confident about the validity of the initial state and the evolution of the resulting structures that we describe below.

\begin{figure*}[t]
\centering
\includegraphics[width=\textwidth]{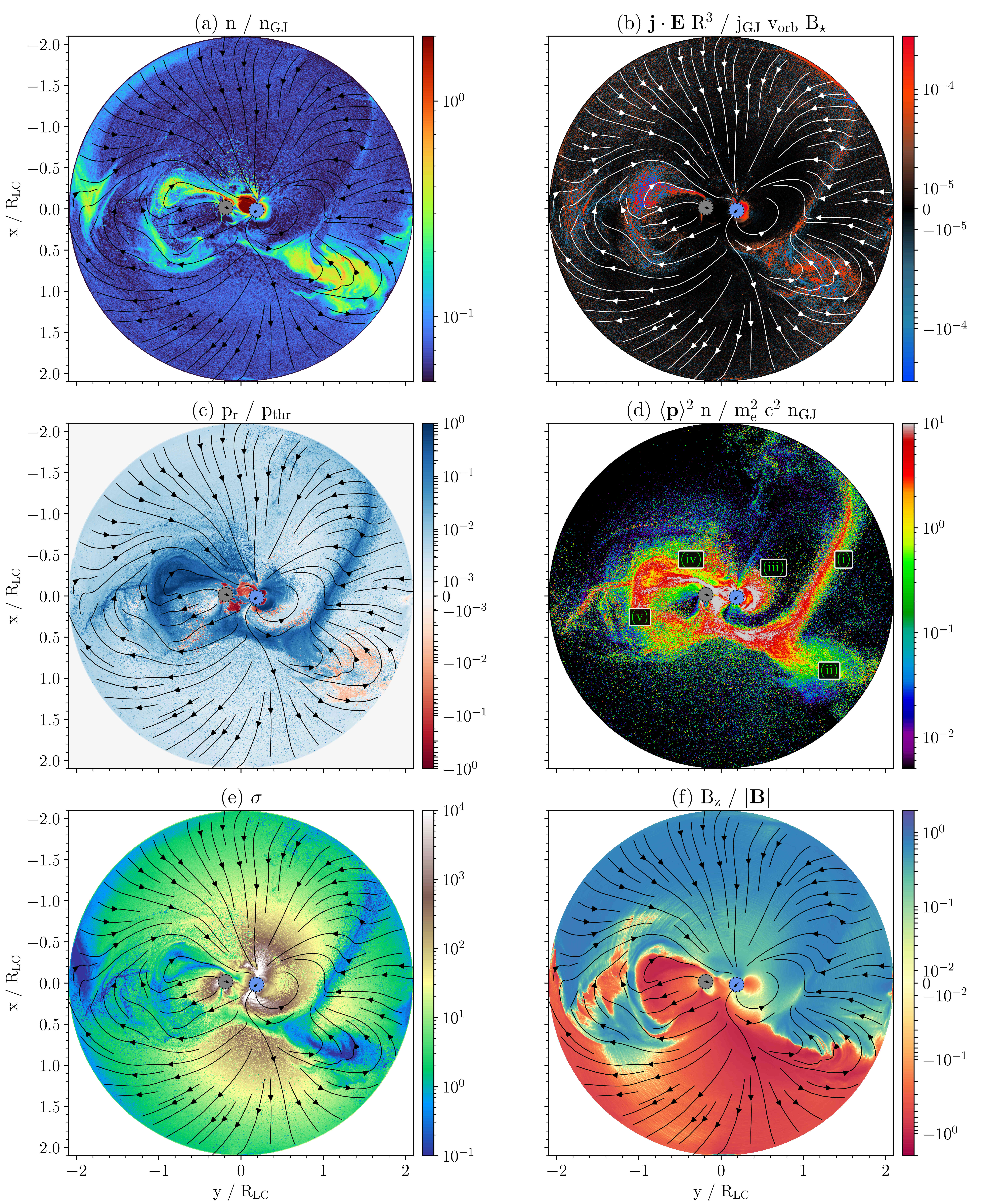}
\caption{
    Orbital-plane cuts of
    (a) electron--positron plasma density,
    (b) electric power density,
    (c) radial component of the bulk momentum normalized to $p_\mathrm{thr}=m_\mathrm{e}c\sqrt{\gamma_\mathrm{thr}^2-1}$,
    (d) second moment of the particle distribution function,
    (e) magnetization $\sigma=B^2/(4\pi\langle\gamma\rangle n m_\mathrm{e}c^2)$, where $\langle\gamma\rangle$ is the distribution-averaged Lorentz factor,
    and (f) the out-of-plane magnetic field component at orbital phase 0.52.
    Quantities are normalized using $\jgj\equiv\rho_\mathrm{GJ}c\equiv\ngj{e c}$, the stellar orbital velocity $\vorb$, and the surface field $\Bstar\equiv\Bone$. 
    Streamlines show the in-plane magnetic field, and the Roman labels in panel~(d) mark the plasma features as in Figure~\ref{fig:render}.
    Star \startwo\, is shown in gray, whereas star \starone\,in blue.\label{fig:cuts}
}
\end{figure*}

The orbital twisting of the magnetic-field lines influences the main magnetospheric properties.
Magnetic-field lines can form closed loops between the stars in the orbital plane that can be cut by the stellar orbital motion.
The resulting behavior at the representative orbital phase at which the investigated features appear simultaneously is shown in a volume rendering in Figure~\ref{fig:render} and in orbital-plane cuts in Figure~\ref{fig:cuts} for the same orbital phase.
Figure~\ref{fig:render} presents the electric current density normalized to the Goldreich--Julian current density, $j_\mathrm{GJ}\equiv{\Omorb\Bone/2\pi}$.
The axis triad marks the orientation of the simulation frame.

A wave emerges when the weak-field star (A') sweeps across a closed magnetic loop connecting opposite magnetic polarities of the stars.
The closed loop is first compressed and filled with relativistically hot plasma attracted from the stellar polar caps or created by pair production.
Once the orbital motion cuts the connecting field lines, the trapped plasma expands rapidly, launches the wave in region (i) (Figures~\ref{fig:render}--\ref{fig:cuts}), and is supported by magnetic pressure of $\Bone$ (region (iii)). 
After analyzing the wave-vector direction and orientation of magnetic and electric-field perturbations, we identify this wave as a fast magnetosonic wave (FMSW).

The FMSW has an initial magnetic-field amplitude comparable to that in closed loops, about $\Btwo$.
As it expands, the background dipolar field decreases with distance faster than the wave amplitude, so that beyond the light cylinder at $\rlc$ the wave becomes nonlinear.
Similar waves are invoked in magnetar magnetospheres \citep{Chen2022,Beloborodov2023,Bernadi2025} as monster shocks and are proposed as FRB emission sources, making the FMSW region a promising site for coherent radiation.

The plasma flares (v) is associated with a magnetically reconnecting current sheet (iv) that appears when field lines of star \startwo\,are forced to approach oppositely oriented field lines of star \starone.
The current sheet starts near the stars, accelerates particles to energies at which they produce new pairs, and eventually develops into a plasma flare, appearing at a distance of about $\rlc$.
Simultaneously, the expanding fireball (iii) is released oppositely to the reconnection because the expansion of related field lines is not constrained by \startwo.
The flare is released as the stellar orbits advance.
After $P_\mathrm{orb}/2$, the events (i)--(v) appear approximately symmetrically to the orbital axis.

\begin{figure*}[t]
\centering
\includegraphics[width=0.98\textwidth]{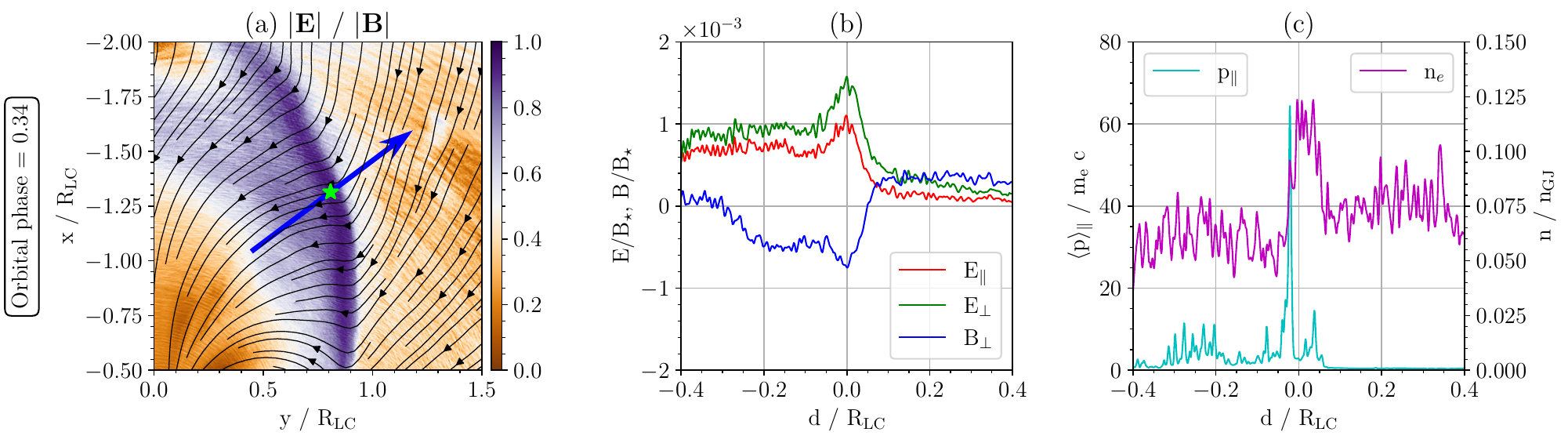} \\
\includegraphics[width=0.98\textwidth]{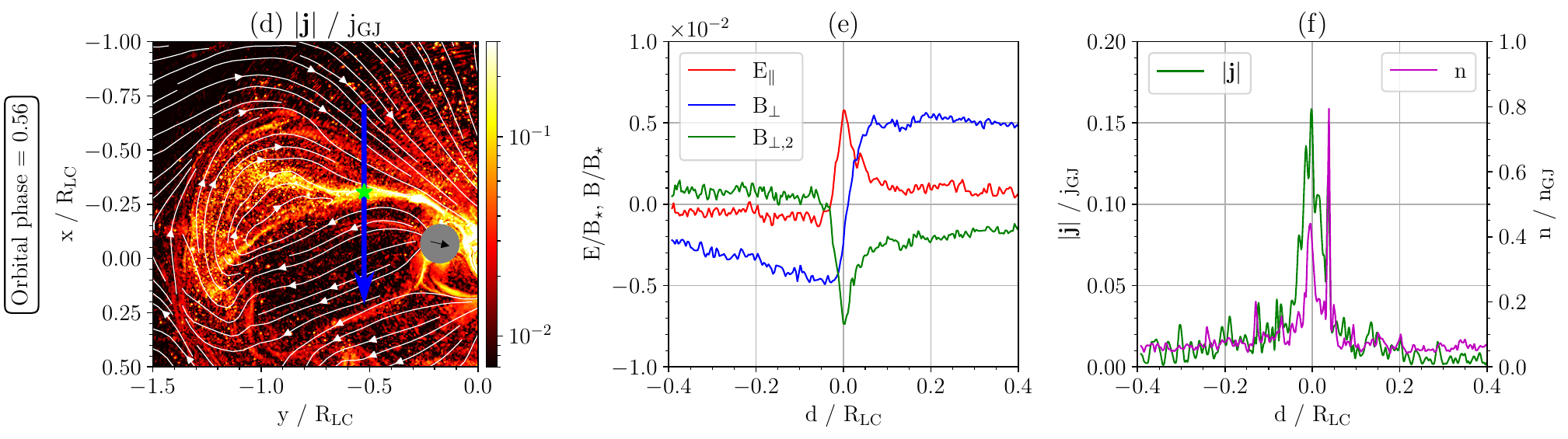}
\caption{Profiles across two structures taken at orbital phases when the structures are clearest.
    \textit{Top:}~the monster shock (i). 
    (a) Electric-to-magnetic field ratio. A 1D cut is denoted by the blue arrow, oriented in the direction of increasing $\boldsymbol{d}$.
    (b) Profiles along $d=|\boldsymbol{d}|$ of the in-plane parallel and perpendicular to $\boldsymbol{d}$ electric field and out-of-plane magnetic field. 
    (c) Parallel bulk momentum and plasma density profiles.
    \textit{Bottom:} the reconnection current sheet (iv) that eventually produces the plasma flare (ii).
    (d) Electric current density.
    (e) In-plane and out-of-plane electric and magnetic field profiles. 
    (f) Total current density and plasma density.
    The green stars mark the wave maximum and current-sheet positions ($d=0$).
    The lines in panels (a) and (d) show the in-plane magnetic field.
\label{fig:cutprofiles}}
\end{figure*}

The internal structure of the FMSW is resolved in Figure~\ref{fig:cutprofiles}(a--c) for a one-dimensional cut taken across the wave. 
The FMSW character is confirmed by its polarization: the wave electric field $\delta\boldsymbol{E}$ is oriented perpendicular to the $\boldsymbol{k}$--$\delta\boldsymbol{B}$ plane, where $\boldsymbol{k}$ is the wave vector and $\delta\boldsymbol{B}$ is the wave amplitude. 
The electric field at the leading edge of the wave ($|\boldsymbol{d}|>0$, where $\boldsymbol{d}$ is the vector along the cut across the region) accelerates the ambient plasma, and a fraction of the particles reach energies sufficient to trigger pair production.
Both pair production and wave compression enhance the plasma density near the perturbation maximum ($|\boldsymbol{d}|\approx0$) and, also due to the increase in average Lorentz factor, reduce $\sigma$ below unity in the FMSW region (Figure~\ref{fig:cuts}(e)).
Moreover, the FMSW propagates outward with radial momentum $p_\mathrm{r}\rightarrow{p}_\mathrm{thr}\equiv\sqrt{\gamma_\mathrm{thr}^2-1}m_\mathrm{e}c$ and high values of $\boldsymbol{j}\cdot\boldsymbol{E}$ and $W_\mathrm{k}$, indicating strong energy transfer from the fields to plasma heating and acceleration to relativistic energies. 
The plasma is advected together with the wave because of its relativistic bulk-flow momentum.

The reconnection signatures are shown in Figure~\ref{fig:cutprofiles}(d--f). 
The in-plane and out-of-plane magnetic-field components reverse at $d=0$, while unequal magnitudes of $B_\mathrm{\perp,2}$ on both sides indicate a slight pressure imbalance, with magnetic pressure advecting the sheet toward smaller $d$. 
The current density and plasma density are strongly enhanced inside the sheet, reflecting a partial balance of magnetic pressure by the plasma pressure.
The plasma density in the sheet is enhanced by pair creation from particles accelerated by the electric field in the current sheet.

\begin{figure*}[t]
\centering
\includegraphics[width=0.35\textwidth]{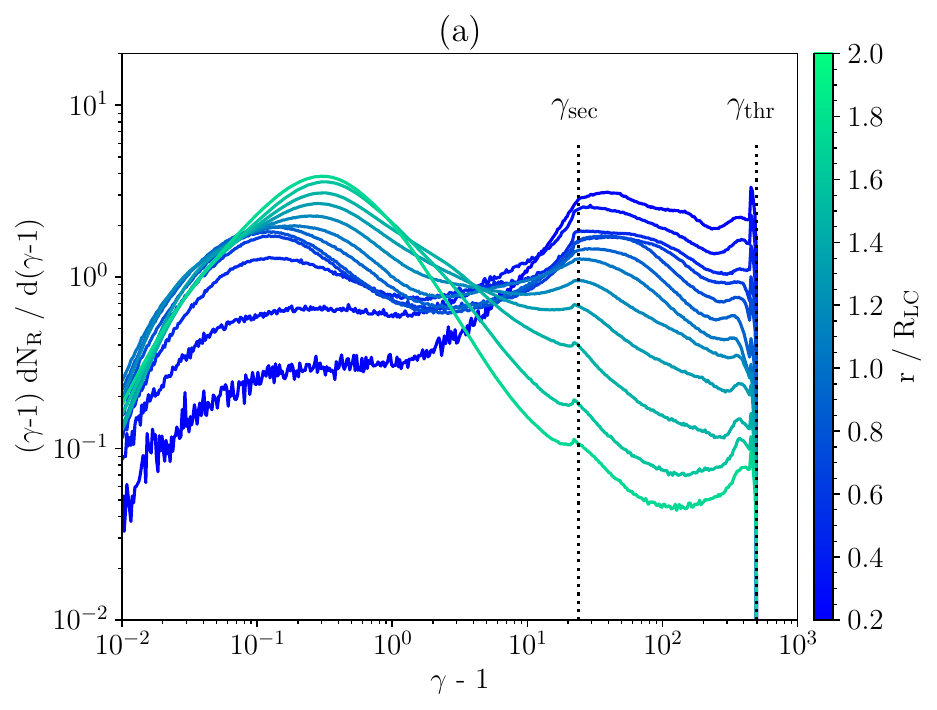}
\includegraphics[width=0.31\textwidth]{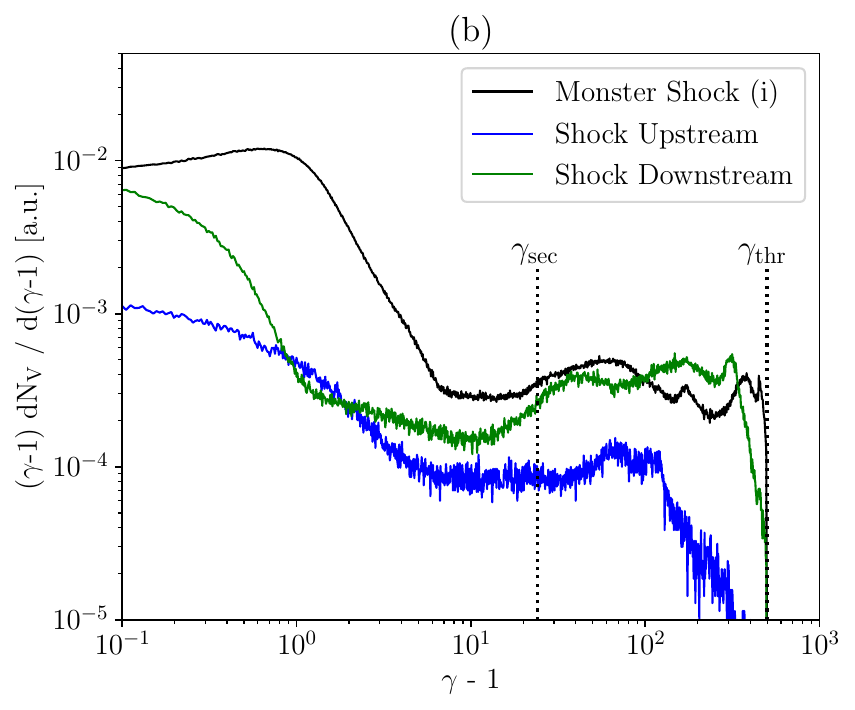}
\includegraphics[width=0.31\textwidth]{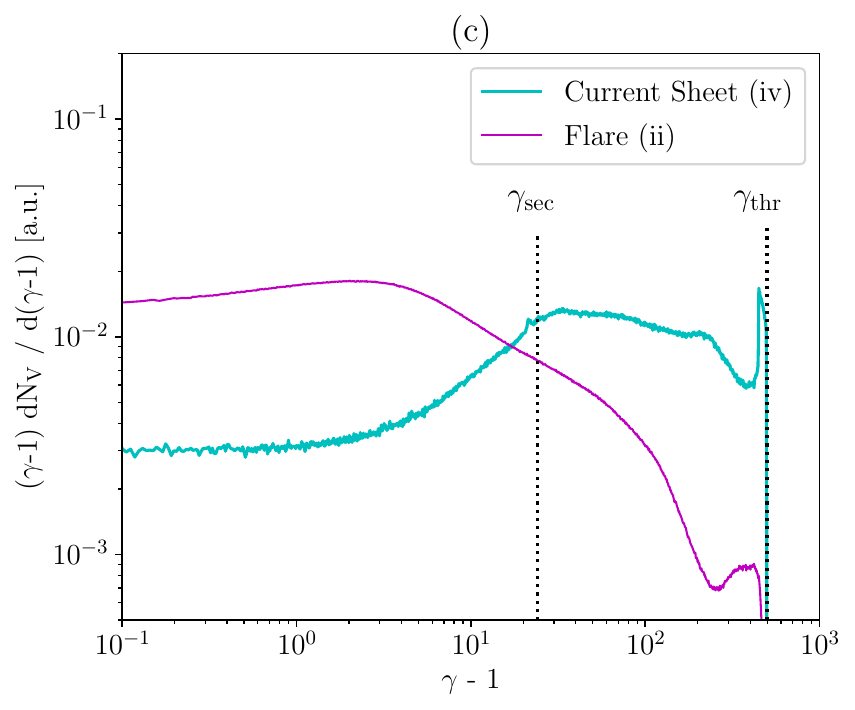}
\caption{
    Particle energy spectra: 
    (a) radial dependence from the center of mass, normalized as $\int_{-\infty}^{\infty}\mathrm{d(\gamma-1)}\left(\mathrm{d}N_\mathrm{r}/\mathrm{d}(\gamma-1)\right)=1$ and averaged over an orbit,
    (b) in the monster shock region, its upstream and downstream, and 
    (c) in the magnetic reconnection current sheet and flare regions.
    Both $N_\mathrm{V}$ distributions are normalized to the average number of PPC in the region and are taken at the orbital phases as in Figure~\ref{fig:cutprofiles}.
    \label{fig:spectrum}
    }
\end{figure*}

\subsection{Particle spectra and Poynting flux}
The particle energy spectra vary strongly with location and time (Figure~\ref{fig:spectrum}).
In general, the spectra are truncated at $\gthr$, above which the primary particles produce pairs, and hence, accumulate just below $\gthr$.
Likewise, injected pairs often enhance the particle number at Lorentz factors near $\gsec$.

The spectra as functions of radius from the center of mass are shown in Figure~\ref{fig:spectrum}(a).
In the inner parts of the magnetosphere, particles accelerated to and beyond $\gsec$ dominate, whereas beyond the light cylinder, $r>\rlc$, the plasma is dominated by colder background plasma.
Nonetheless, the spectral tail still reaches $\gthr$.

While the plasma upstream of the FMSW (Figure~\ref{fig:spectrum}b) is relatively dilute, and particles typically do not reach $\gthr$, inside the FMSW particles are accelerated to $\gamma_\mathrm{thr}$, and plasma density is enhanced.
Downstream of the FMSW, the high-energy part of the particle distribution remains enhanced.
Magnetic reconnection very efficiently accelerates particles to $\gamma_\mathrm{thr}$, and the distribution contains only a weak background plasma component. But once the flare is formed, the high-energy part of the distribution falls off, as the accelerated particles from the current sheet are dispersed in the much larger volume of the flare region.

\begin{figure}[t]
\centering
\includegraphics[width=\columnwidth]{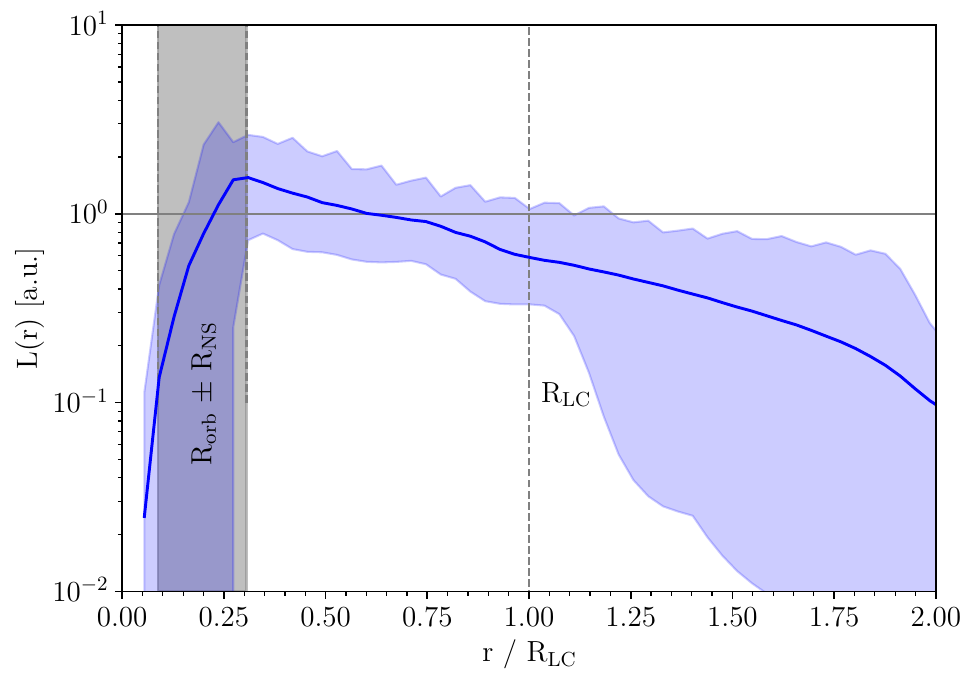}
\caption{
    Radial dependence of the average Poynting flux luminosity from the center of mass (blue line).
    The blue-shaded region denotes a range of luminosities obtained from several time steps during one orbit.
    The gray-shaded region indicates the stellar distances.
    \label{fig:Poynting}
    }
\end{figure}
The radial Poynting-flux luminosity as a function of distance from the center of mass is shown in Figure~\ref{fig:Poynting}.
We define the luminosity as $L(r)=\oint_{\Sigma}{\boldsymbol{S}(r)}\cdot\mathrm{d}\boldsymbol{\Sigma}$, where $\boldsymbol{S}=\frac{c}{4\pi}\boldsymbol{E}\times\boldsymbol{B}$ is the Poynting flux where $\boldsymbol{\Sigma}$ is the sphere radial unit vector.
The average luminosity decreases for larger orbital radii, reaching only about 10\,\% of its maximum at a distance of $2\rlc$.
This is faster than in pulsar winds outside the (stellar-spin) light cylinder \citep{Cerutti2017}.
Moreover, the luminosity varies strongly during the orbit, especially outside $\rlc$, where it can fall to 1\,\% of the maximum at $1.6\rlc$ for a certain orbital phase.
This indicates that we can expect very large variation of electromagnetic luminosities of the BNS, reaching about ${\Delta}L_\mathrm{orb}\approx\Bstar^2R_\mathrm{NS}^6\Omorb^6D^2/2c^5\approx3\times10^{40}$\,erg\,s$^{-1}$, where $D=(GM/\Omorb^2)^{\frac{1}{3}}$ is the stellar orbital radius, where we assume $\Bstar\equiv10^{12}$\,G \citep{Wada2020}.
The spin-down luminosities of the stars $L_\mathrm{A}\approx10^{35}$\,erg\,s$^{-1}$ and $L_\mathrm{B}\approx10^{37}$\,erg\,s$^{-1}$ remain smaller than the orbital Poynting flux luminosity.

\subsection{FRB Energy Sources and Emission Regions}
The monster shock is a candidate region for FRB emission because the FMSW can produce soliton-like oscillations and be converted into an X-mode wave that can escape the plasma \citep{Vanthieghem2025}.
Although we detect certain regular structures in the plasma charge density and the electric field of the wave, we cannot confirm whether these are soliton-like waves appearing at kinetic scales.
Furthermore, we estimate the FMSW energy as the energy that is eventually converted into coherent radiation.
For this estimate, we fix an energy-conversion factor from the FMSW to the coherent radiation as $\xi\equiv10^{-3}$ \citep{Vanthieghem2025}.
We find that most of the shock energy is contained in the electromagnetic fields.
For a wave amplitude of $1.5\times10^{-3}\Bstar$ and a volume $(0.1\times1.5\times1.5)\rlc^3$, the monster shock contains $6\times10^{37}$\,erg in energy.
If that is emitted into a solid angle of $1$\,sr in a time interval $0.1\rlc/c\approx38\,\mathrm{{\mu}s}$, we obtain an upper limit on the isotropic coherent radio luminosity of $2\times10^{40}$\,erg\,s$^{-1}$.
Similarly, we estimate the emitted power from the Poynting flux at the monster shock as $4\times10^{42}$\,erg\,s$^{-1}$, implying a radio luminosity of $4\times10^{39}$\,erg\,s$^{-1}$.
The former estimate is slightly above the typical FRB luminosity range of $10^{38}$--$10^{42}$\,erg\,s$^{-1}$, indicating that the FMSW could have a smaller amplitude (for example, BNS emits FRBs at an earlier time before the merger), or $\xi$ could be smaller than $10^{-3}$.

The flare region is another candidate for FRB emission.
Its electromagnetic fields could provide a power of $4\times10^{44}$\,erg\,s$^{-1}$ and an isotropic radio luminosity of $6\times10^{42}$\,erg\,s$^{-1}$, emitted during a time interval of $\Porb/2\approx1.2$\,ms.
The Poynting flux luminosity is slightly lower, at $3\times10^{40}$\,erg\,s$^{-1}$, with an isotropic radio luminosity of $4\times10^{39}$\,erg\,s$^{-1}$.
Nonetheless, further analysis shows that the plasma in the flare is in a turbulent state.
Turbulence spanning several spatio-temporal orders of magnitude ($\Delta\nu/\nu\gg1$) can degrade the narrowband character ($\Delta\nu/\nu\lesssim1$) of the emitted radiation, where $\Delta\nu$ is the range of emitted FRB frequencies and $\nu$ is the central frequency.

The plasma kinetic energy in the shock and flare regions is a few orders of magnitude smaller than the electromagnetic field energy.
Without an additional energy supply, the kinetic energy is therefore not sufficiently high to supply the energy budget required for an FRB.
Hence, an energy supply from the fields is necessary.

\subsection{Dependence on the Binary Configuration}\label{sec:param}
To isolate the role of system parameters, we ran a sequence of simpler magnetospheric configurations, varying the ratio of the stellar spin periods, the ratio
of the stellar magnetic fields, and the angle between the magnetic and rotational axes with the following approximate trends.
Once $\Bone\gg\Btwo$, the flaring activity is confined to a smaller solid angle behind the star \startwo. 
For $\Porb<\Ptwo$ and $\Pone$, the spin of star~\starone\,governs the \emph{direction} in which flares are ejected because of its strong magnetic field, whereas the spin of star~\startwo\,\emph{modulates} the intensity of all events. 
The topology dependence is quantitatively consistent with the force-free results of \citet{MostPhilippov2022} and \citet{Skiathas2025}, who likewise found the dissipation and emission geometry is set by the relative orientation and strengths of the stellar dipoles.

\section{Discussion and Conclusions} \label{sec:disc}
We present the first kinetic model of a BNS system in which orbital motion dominates over stellar spin in its impact on the magnetospheric dynamics.
Our configuration is inspired by the double pulsar \PSR\,evolved to its late inspiral phase before the merger.

The simulation identifies the mechanisms that control the system.
First, the ratio of the stellar magnetic field intensities governs the emission geometry of the flare, with the flare becoming more focused at higher ratios because the star with the stronger magnetic field dominates the common magnetosphere.
Second, the spin of the strong-field star sets the direction along which the monster shock, flare, and fireball are ejected, while the spin of the weak-field star modulates the amplitudes of these events as the magnetic field lines of both stars interact.

During the orbital motion, we identify four regions of strong electromagnetic energy release, occurring twice per orbital period as long as $\Pone\gg\Ptwo>\Porb$:
(i) an FMSW forming a monster shock, (ii) a plasma flare, (iii) an expanding fireball, and (iv) magnetic reconnection that powers the flare.
These regions contain sufficient energy to explain FRB luminosities, but the conversion efficiency and mechanisms by which global electromagnetic field energy converts into narrowband coherent emission require detailed investigation at plasma kinetic scales.

The monster shock is a promising coherent radio source for FRBs because its large-amplitude FMSWs can excite coherent radio emission, similar to the case of magnetar flares \citep{Lyubarsky2020,Yuan2020,Vanthieghem2025,Solanki2026}.
The FMSW obtained in the simulation can reach amplitudes greater than the background magnetic field beyond $1\,\rlc$. 
However, whether this energy emerges as an observable burst depends on the escape of the radiation from the magnetosphere \citep{Beloborodov2021,Lyutikov2024}, which remains to be quantified. 

The evaluation of flare FRB emission is constrained by the strongly turbulent nature of the flare, whereas the FRB signal is narrowband. 
Also, the magnetic reconnection within the flare is expected to be a powerful radiator \citep{Philippov2019}, but the waves produced by reconnection are broadband.
Moreover, the expanding fireball could radiate coherently if the mixing of the plasma species induces density cavities that modulate the emission, in analogy with the magnetar-fireball mechanism\footnote{Araudo, A., Benáček, J., Vieyro, F., Romero, G.E., \& Tikhonchuk, V., in preparation.}.

Although the BNS inspiral scenario is often considered to produce a single ``one-off'' FRB, BNS system can produce many FRBs before the merger.
This repetition decreases the size of the BNS inspiral population in the Universe required to explain the one-off FRB detection rate.

Our identification of the FMSW as the coherent channel complements the recent kinetic study of \citet{Parsons2026}, who, for the stellar spin-dominated regime, attributed FRB-like radio transients to plasmoid mergers in the trailing current sheet. 
In our orbiting configuration, the coherent channel could be the FMSW launched when orbital motion cuts magnetic loops.
The FMSW can propagate along open field lines and beyond the light cylinder; its later evolution remains to be quantified.
A kinetic identification of the potential coherent-emission region in pre-merger magnetospheres connects the physics of FRBs to the multimessenger astronomy of compact-binary coalescence, providing targets for searches for radio precursors of gravitational-wave events.

In summary, an orbiting kinetic plasma model at a BNS at fixed late-inspiral separation reveals plasma events in which coherent radio emission could be produced, with the FMSW as the candidate in a late-inspiral double pulsar configuration. 
For future work, we suggest focusing on individual emission regions in detail to investigate plasma kinetic properties and their relation to the emission of coherent radiation and its escape.

\begin{acknowledgments}
J.B. thanks A. Vanthieghem for a discussion about FMSW conversion.
We acknowledge the developers of the ACRONYM code (Verein zur F\"orderung kinetischer Plasmasimulationen e.V.).
J.B. also acknowledges the support by the German Science Foundation (DFG) project BE~7886/2-1. 
This project has received funding from the European Union’s Horizon Europe research and innovation programme under the Marie Skłodowska-Curie grant agreement No 101203963.
T.D.\ acknowledge funding from the EU Horizon under ERC Starting Grant, no.\ SMArt-101076369. 
Views and opinions expressed are those of the authors only and do not necessarily reflect those of the European Union or the European Research Council.
Neither the European Union nor the granting authority can be held responsible for them.
The authors gratefully acknowledge the Gauss Centre for Supercomputing e.V. (\url{www.gauss-centre.eu}) for partially funding this project by providing computing time on the GCS Supercomputer SuperMUC-NG at Leibniz Supercomputing Centre (www.lrz.de), projects pn73ne and pn52ku.
\end{acknowledgments}



\appendix

\section{Details of the BNS numerical setup} \label{appendixA}

We performed the simulation with the explicit, fully kinetic, special-relativistic PIC code \textsc{Acronym} \citep{Kilian2012,Benacek2026} on a Yee lattice with a modified fourth-order finite-difference time-domain field solver \citep{Greenwood2004}, particle advancement with the Vay pusher, and second-order shape functions, appropriate for relativistic pair plasmas. 

The electromagnetic fields are evolved on a cubic domain of $1200^3$ grid cells, with the center of mass of the binary placed in the center of the
domain. 
The NS radius of $\rns=12$~km is resolved by $30$ grid cells.
We evolve the system for a few orbital periods, corresponding to $20{,}000$ time steps, until the initial transition effects disappear. One orbit is resolved by about $4{,}500$ time steps.

We use flat spacetime and Newtonian orbits in the inertial reference frame of the center of mass.
The corresponding Keplerian separation is around $3.6\,\rns$, which we keep fixed because the inspiral motion does not significantly change the separation during a few orbits. 

At the inner boundaries, located on the stellar surfaces, we impose a dipole field transformed into the orbiting reference frame of each star and inject pair plasma in a comoving frame with a small thermal velocity to keep $20$ particles (electrons plus positrons) per cell (PPC), corresponding to $n\approx2\ngj$, where
$\ngj=\Omorb\Bone/(2\pi c e)$ is the nominal orbital normalization of the Goldreich--Julian (GJ) number density \citep{Goldreich1969}, with $e$ the elementary charge. 

The outer boundary is an absorbing sphere that removes all particles reaching a radius of $596{\Delta}x\approx2.08\rlc$ from the center and gradually absorbs electromagnetic waves within a layer of four grid cells.
We limit the number of particles per cell (PPC) to a maximum of 150, allowing the plasma to reach $\sigma<1$, and employ a $\mathrm{div}\boldsymbol{B}$ projection cleaning method \citep{Brackbill1980} to minimize any inconsistencies that can appear in the electromagnetic fields as the stars move through the grid.

New pairs are injected when particles reach a Lorentz factor $\gamma_\mathrm{thr}=500$, with ten percent of the particle energy removed and equally distributed among the newly created particles, each having a Lorentz factor $\gsec\approx24$.


\begin{thebibliography}{}

\bibitem[Abac et al.(2024)]{Abac2024} Abac, A., Dietrich, T., Buonanno, A., Steinhoff, J., \& Ujevic, M.\ 2024, \prd, 109, 024062

\bibitem[Abbott et al.(2017)]{Abbott2017} Abbott, B. P., Abbott, R., Abbott, T. D., et al.\ 2017, \prl, 119, 161101

\bibitem[Acernese et al.(2015)]{Acernese2015} Acernese, F., Agathos, M., Agatsuma, K., et al.\ 2015, CQG, 32, 024001

\bibitem[Araudo et al.(2026)]{Araudo2026} Araudo, A., et al. 2026, in preparation

\bibitem[Beloborodov(2021)]{Beloborodov2021} Beloborodov, A.~M.\ 2021, \apjl, 922, 1, L7

\bibitem[Beloborodov(2023)]{Beloborodov2023} Beloborodov, A.~M. 2023, \apj, 959, 1, 34

\bibitem[Ben{\'a}{\v{c}}ek et al.(2026)]{Benacek2026} Ben{\'a}{\v{c}}ek, J., Jessner, A., Pohl, M., et al. 2026, \aap, 707, A316 

\bibitem[Bernardi et al.(2025)]{Bernadi2025} Bernardi, D., Yuan, Y., \& Chen, A.~Y.\ 2025, \prl, 135, 26, 265201

\bibitem[Brackbill \& Barnes(1980)]{Brackbill1980} Brackbill, J.~U. \& Barnes, D.~C.\ 1980, Journal of Computational Physics, 35, 3, 426

\bibitem[Burgay et al.(2003)]{Burgay2003} Burgay, M., D'Amico, N., Possenti, A., et al. 2003, \nat, 426, 531

\bibitem[Carrasco \& Shibata(2020)]{Carrasco2020} Carrasco, F., \& Shibata, M. 2020, \prd, 101, 063017

\bibitem[Cerutti et al.(2015)]{Cerutti2015} Cerutti, B., Philippov, A., Parfrey, K., \& Spitkovsky, A. 2015, \mnras, 448, 606

\bibitem[Cerutti \& Philippov(2017)]{Cerutti2017} Cerutti, B., \& Philippov, A.~A. 2017, \aap, 607, A134

\bibitem[Chen et al.(2022)]{Chen2022} Chen, A.~Y., Yuan, Y., Li, X., et al.\ 2022, arXiv:2210.13506

\bibitem[Cordes \& Chatterjee(2019)]{Cordes2019} Cordes, J.~M., \& Chatterjee, S. 2019, \araa, 57, 417

\bibitem[Crinquand et al.(2019)]{Crinquand2019} Crinquand, B., Cerutti, B., \& Dubus, G.\ 2019, \aap, 622, A161

\bibitem[Dietrich et al.(2020)]{Dietrich2020} Dietrich, T., Coughlin, M. W., Pang, P. T. H., Bulla, M., Heinzel, J., Issa, L., Tews, I., \& Antier, S.\ 2020, Science, 370, 1450

\bibitem[Goldreich \& Julian(1969)]{Goldreich1969} Goldreich, P., \& Julian, W.~H. 1969, \apj, 157, 869

\bibitem[Greenwood et al.(2004)]{Greenwood2004} Greenwood, A.~D., Cartwright, K.~L., Luginsland, J.~W., et al.\ 2004, Journal of Computational Physics, 201, 2, 665. doi:10.1016/j.jcp.2004.06.021

\bibitem[Kilian et al.(2012)]{Kilian2012} Kilian, P., Burkart, T., \& Spanier, F. 2012, in High Performance Computing in Science and Engineering'11, ed. W.~E. Nagel, D.~B. Kr\"oner, \& M.~M. Resch (Berlin: Springer), 5

\bibitem[Koehn et al.(2025)]{Koehn2025} Koehn, H., et al.\ 2025, PRX, 15, 021014

\bibitem[Kramer et al.(2006)]{Kramer2006} Kramer, M., Stairs, I.~H., Manchester, R.~N., et al. 2006, Science, 314, 97

\bibitem[Kramer et al.(2021)]{Kramer2021} Kramer, M., Stairs, I.~H., Manchester, R.~N., et al. 2021, Physical Review X, 11, 041050

\bibitem[LIGO Scientific Collaboration et al.(2015)]{LIGO2015} LIGO Scientific Collaboration, Aasi, J., Abbott, B. P., et al.\ 2015, QCG, 32, 074001

\bibitem[Lyne et al.(2004)]{Lyne2004} Lyne, A.~G., Burgay, M., Kramer, M., et al. 2004, Science, 303, 1153

\bibitem[Lyubarsky(2020)]{Lyubarsky2020} Lyubarsky, Y. 2020, \apj, 897, 1

\bibitem[Lyutikov(2024)]{Lyutikov2024} Lyutikov, M. 2024, \mnras, 529, 2180

\bibitem[Lyutikov(2026)]{Lyutikov2026} Lyutikov, M.\ 2026, \apjl, 1004, 2, L43

\bibitem[Mahlmann \& Beloborodov(2025)]{Mahlmann2025} Mahlmann, J.~F., \& Beloborodov, A.~M. 2025, \apjl, 981, L17

\bibitem[Most \& Philippov(2020)]{MostPhilippov2020} Most, E.~R., \& Philippov, A.~A. 2020, \apjl, 893, L6

\bibitem[Most \& Philippov(2022)]{MostPhilippov2022} Most, E.~R., \& Philippov, A.~A. 2022, \mnras, 515, 2710

\bibitem[Most \& Philippov(2023)]{MostPhilippov2023} Most, E.~R., \& Philippov, A.~A. 2023, \prl, 130, 245201

\bibitem[Palenzuela et al.(2013)]{Palenzuela2013} Palenzuela, C., Lehner, L., Ponce, M., et al. 2013, \prl, 111, 061105

\bibitem[Parsons et al.(2026)]{Parsons2026} Parsons, J., Spitkovsky, A., Philippov, A., et al.\ 2026, \apjl, 1007, 2, L50. doi:10.3847/2041-8213/ae9088

\bibitem[Petroff et al.(2019)]{Petroff2019} Petroff, E., Hessels, J.~W.~T., \& Lorimer, D.~R. 2019, \aapr, 27, 4

\bibitem[Philippov \& Spitkovsky(2014)]{PhilippovSpitkovsky2014} Philippov, A.~A., \& Spitkovsky, A. 2014, \apjl, 785, L33

\bibitem[Philippov et al.(2019)]{Philippov2019} Philippov, A., Uzdensky, D.~A., Spitkovsky, A., et al.\ 2019, \apjl, 876, 1, L6

\bibitem[Skiathas et al.(2025)]{Skiathas2025} Skiathas, D., Kalapotharakos, C., Wadiasingh, Z., et al.\ 2025, \apj, 994, 1, 131

\bibitem[Solanki et al.(2026)]{Solanki2026} Solanki, S., Mahlmann, J., \& Philippov, A.\ 2026, arXiv:2606.19450

\bibitem[Yuan et al.(2020)]{Yuan2020} Yuan, Y., Beloborodov, A.~M., Chen, A.~Y., \& Levin, Y. 2020, \apjl, 900, L21

\bibitem[Vanthieghem \& Levinson(2025)]{Vanthieghem2025} Vanthieghem, A. \& Levinson, A.\ 2025, \prl, 134, 3, 035201

\bibitem[Wada et al.(2020)]{Wada2020} Wada, T., Shibata, M., \& Ioka, K.\ 2020, Progress of Theoretical and Experimental Physics, 2020, 10, 103E01

\end{thebibliography}
\end{document}